\documentclass[12pt]{article}

\usepackage{amsmath,amsfonts,amssymb,color,epsfig,graphicx,latexsym,revsymb,theorem,url,verbatim}

\usepackage{epstopdf}

\usepackage{caption}

\usepackage[cm]{fullpage}

\newtheorem{definition}{Definition}
\newtheorem{proposition}[definition]{Proposition}
\newtheorem{lemma}[definition]{Lemma}

\newtheorem{theorem}[definition]{Theorem}
\newtheorem{corollary}[definition]{Corollary}
\newtheorem{conjecture}[definition]{Conjecture}

\newtheorem{remark}[definition]{Remark}
\newtheorem{example}[definition]{Example}
\newtheorem{question}[definition]{Question}

\def\squareforqed{\hbox{\rlap{$\sqcap$}$\sqcup$}}
\def\qed{\ifmmode\squareforqed\else{\unskip\nobreak\hfil
\penalty50\hskip1em\null\nobreak\hfil\squareforqed
\parfillskip=0pt\finalhyphendemerits=0\endgraf}\fi}
\def\endenv{\ifmmode\;\else{\unskip\nobreak\hfil
\penalty50\hskip1em\null\nobreak\hfil\;
\parfillskip=0pt\finalhyphendemerits=0\endgraf}\fi}
\newenvironment{proof}{\noindent \textbf{{Proof.~} }}{\qed}
\def\Dbar{\leavevmode\lower.6ex\hbox to 0pt
{\hskip-.23ex\accent"16\hss}D}
\makeatletter
\def\url@leostyle{%
  \@ifundefined{selectfont}{\def\UrlFont{\sf}}{\def\UrlFont{\small\ttfamily}}}
\makeatother
\def\bcj{\begin{conjecture}}
\def\ecj{\end{conjecture}}
\def\bcr{\begin{corollary}}
\def\ecr{\end{corollary}}
\def\bd{\begin{definition}}
\def\ed{\end{definition}}
\def\bea{\begin{eqnarray}}
\def\eea{\end{eqnarray}}
\def\bem{\begin{enumerate}}
\def\eem{\end{enumerate}}
\def\bex{\begin{example}}
\def\eex{\end{example}}
\def\bim{\begin{itemize}}
\def\eim{\end{itemize}}
\def\bl{\begin{lemma}}
\def\el{\end{lemma}}
\def\bpf{\begin{proof}}
\def\epf{\end{proof}}
\def\bpp{\begin{proposition}}
\def\epp{\end{proposition}}
\def\bqu{\begin{question}}
\def\equ{\end{question}}
\def\br{\begin{remark}}
\def\er{\end{remark}}
\def\bt{\begin{theorem}}
\def\et{\end{theorem}}

\def\btb{\begin{tabular}}
\def\etb{\end{tabular}}

\newcommand{\nc}{\newcommand}

\def\a{\alpha}
\def\b{\beta}
\def\g{\gamma}
\def\d{\delta}

\def\s{\sigma}

\def\ps{\psi}

\def\G{\Gamma}

\def\Ps{\Psi}

 \nc{\bA}{{\bf A}} \nc{\bB}{{\bf B}} \nc{\bC}{{\bf C}}
 \nc{\bD}{{\bf D}} \nc{\bE}{{\bf E}} \nc{\bF}{{\bf F}}
 \nc{\bG}{{\bf G}} \nc{\bH}{{\bf H}} \nc{\bI}{{\bf I}}
 \nc{\bJ}{{\bf J}} \nc{\bK}{{\bf K}} \nc{\bL}{{\bf L}}
 \nc{\bM}{{\bf M}} \nc{\bN}{{\bf N}} \nc{\bO}{{\bf O}}
 \nc{\bP}{{\bf P}} \nc{\bQ}{{\bf Q}} \nc{\bR}{{\bf R}}
 \nc{\bS}{{\bf S}} \nc{\bT}{{\bf T}} \nc{\bU}{{\bf U}}
 \nc{\bV}{{\bf V}} \nc{\bW}{{\bf W}} \nc{\bX}{{\bf X}}
 \nc{\bZ}{{\bf Z}}

\nc{\cA}{{\cal A}} \nc{\cB}{{\cal B}} \nc{\cC}{{\cal C}}
\nc{\cD}{{\cal D}} \nc{\cE}{{\cal E}} \nc{\cF}{{\cal F}}
\nc{\cG}{{\cal G}} \nc{\cH}{{\cal H}} \nc{\cI}{{\cal I}}
\nc{\cJ}{{\cal J}} \nc{\cK}{{\cal K}} \nc{\cL}{{\cal L}}
\nc{\cM}{{\cal M}} \nc{\cN}{{\cal N}} \nc{\cO}{{\cal O}}
\nc{\cP}{{\cal P}} \nc{\cQ}{{\cal Q}} \nc{\cR}{{\cal R}}
\nc{\cS}{{\cal S}} \nc{\cT}{{\cal T}} \nc{\cU}{{\cal U}}
\nc{\cV}{{\cal V}} \nc{\cW}{{\cal W}} \nc{\cX}{{\cal X}}
\nc{\cZ}{{\cal Z}}

\nc{\hA}{{\hat{A}}} \nc{\hB}{{\hat{B}}} \nc{\hC}{{\hat{C}}}
\nc{\hD}{{\hat{D}}} \nc{\hE}{{\hat{E}}} \nc{\hF}{{\hat{F}}}
\nc{\hG}{{\hat{G}}} \nc{\hH}{{\hat{H}}} \nc{\hI}{{\hat{I}}}
\nc{\hJ}{{\hat{J}}} \nc{\hK}{{\hat{K}}} \nc{\hL}{{\hat{L}}}
\nc{\hM}{{\hat{M}}} \nc{\hN}{{\hat{N}}} \nc{\hO}{{\hat{O}}}
\nc{\hP}{{\hat{P}}} \nc{\hR}{{\hat{R}}} \nc{\hS}{{\hat{S}}}
\nc{\hT}{{\hat{T}}} \nc{\hU}{{\hat{U}}} \nc{\hV}{{\hat{V}}}
\nc{\hW}{{\hat{W}}} \nc{\hX}{{\hat{X}}} \nc{\hZ}{{\hat{Z}}}

\nc{\hn}{{\hat{n}}}

\def\dim{\mathop{\rm Dim}}

\def\lin{\mathop{\rm span}}

\def\min{\mathop{\rm min}}

\def\rank{\mathop{\rm rank}}

\def\w{\mathop{\rm W}}

\def\GL{{\mbox{\rm GL}}}

\def\Un{{\mbox{\rm U}}}

\def\dg{\dagger}

\def\op{\oplus}
\def\ox{\otimes}

\def\ra{\rightarrow}

\newcommand{\bra}[1]{\langle#1|}
\newcommand{\ket}[1]{|#1\rangle}
\newcommand{\proj}[1]{| #1\rangle\!\langle #1 |}
\newcommand{\ketbra}[2]{|#1\rangle\!\langle#2|}
\newcommand{\braket}[2]{\langle#1|#2\rangle}

\newcommand{\abs}[1]{|#1|}

\newcommand{\red}{\textcolor{red}}

\def\Dbar{\leavevmode\lower.6ex\hbox to 0pt
{\hskip-.23ex\accent"16\hss}D}

\begin{document}
\title{Nonlocal and controlled unitary operators of Schmidt rank three}

\author{ Lin Chen $^{1}$ and Li Yu $^{1}$\\
\small ${}^{1}$ Singapore University of Technology and Design, 20
Dover Drive, Singapore 138682 \\
}

\date{\today}

%
\maketitle
\abstract
{Implementing nonlocal unitary operators is an important
and hard question in quantum computing and cryptography. We show
that any bipartite nonlocal unitary operator of Schmidt rank three on the $(d_A
\times d_B)$-dimensional system is locally equivalent to a controlled unitary
when $d_A$ is at most three. This operator can be locally implemented
assisted by a maximally entangled state of Schmidt rank $r=\min\{d_A^2,d_B\}$. We further show that stochastic-equivalent
nonlocal unitary operators are indeed locally equivalent, and
propose a sufficient condition on which nonlocal and controlled
unitary operators are locally equivalent. We also provide the
solution to a special case of a conjecture on the ranks of
multipartite quantum states.}


\section{Introduction}

Implementing multipartite unitary operators is a fundamental task in
quantum information theory. The operators are called \textit{local}
when they are the tensor product of unitary operators locally acting
on subsystems, i.e., they have \textit{Schmidt rank} one. Otherwise
they are called \textit{nonlocal}. It is known that the local
unitary can be implemented by local operations and classical
communication (LOCC) with probability one. Recent research has been
devoted to the decomposition of local unitaries into elementary
operations \cite{kmm13}, and the local equivalence between
multipartite quantum states of fermionic systems under local
unitaries \cite{cdgz13,ccd13,cdg13}.

Nonlocal unitary operators have a more complex structure and play a
more powerful role than local unitaries in quantum computing,
cryptography, and so on. Nonlocal unitaries can create quantum
entanglement between distributed parties \cite{ejp00}, and their
equivalence has been studied under LOCC \cite{dc02}. Nonlocal
unitaries cannot be implemented by LOCC only, even if the
probability is allowed to be close to zero \cite{pv98}.  The understanding of the forms and
implementation schemes of nonlocal unitary operators is still far from complete.  The simplest type of nonlocal unitary
is the \textit{controlled unitary} gates, which are of the general form
$U=\sum_{j=1}^m P_j \ox V_j$ acting on a bipartite Hilbert space $\cH_A \ox \cH_B$, where $P_j$ are orthogonal projectors on $\cH_A$ and $V_j$ are unitaries on $\cH_B$. They can be implemented by a simple nonlocal protocol \cite{ygc10} using a maximally entangled state of Schmidt rank $m$. In this sense the implementation of controlled unitaries is operational. Some other types of nonlocal unitaries are discussed in \cite{ygc10}, but in this paper we will focus on controlled unitaries.  (Note that entirely different implementations are possible if the systems are deemed to be near enough so as to allow direct quantum interactions between them, e.g. multiqubit controlled gates can be decomposed into certain elementary gates \cite{bbc95}). Recently an interesting connection between nonlocal and controlled unitaries was found: they are
locally equivalent when they have Schmidt rank 2 \cite{cy13}.  In this case their implementations are the same. So
it is important to strengthen this connection for operationally implementing more nonlocal unitaries.


\begin{figure}[ht]
\includegraphics{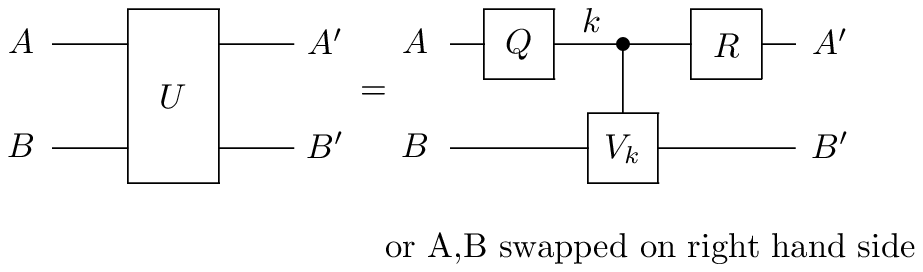}
\captionsetup{font=scriptsize,width=0.8\textwidth}
\caption{Any bipartite unitary $U$ on $d_A \times d_B$ system of Schmidt rank 3 is locally
equivalent to a controlled unitary when $d_A=2,3$, where the controlling side may be $A$ or $B$.  This is expressed as $U=(Q\ox I)(\sum^{d_A}_{k=1}\proj{k}\ox V_k)(R\ox I)$ or $U=(I\ox Q)(\sum^{d_B}_{k=1}V_k \ox \proj{k})(I\ox R)$, where $V_k$, $Q$ and $R$ are local unitaries. The output systems $A'$ and $B'$ are assumed to be of the same size as $A$ and $B$, respectively.}
\label{fig1}
\end{figure}

In this paper we show that any bipartite unitary operation $U$ of Schmidt rank 3 on $d_A \times d_B$ system is locally
equivalent to a controlled unitary when $d_A=2,3$, see Theorems
\ref{thm:rk3da2} and \ref{thm:3xd}. This is illustrated in Fig.~\ref{fig1}.
They not only imply the method of implementing $U$ but also simplify the structure of $U$. We also
propose an operational method of explicitly decomposing $U$ into the
form of controlled unitaries in the end of Sec. \ref{sec:conj}. As
an application we can simplify the problem of deciding the
stochastic local equivalence, namely SL-equivalence of two bipartite unitaries of Schmidt rank 3 with
$d_A=2,3$. This is based on Theorem \ref{thm:sllu} that any two
SL-equivalent nonlocal unitary operators are locally equivalent.
Using this theorem we provide a sufficient condition by which a
bipartite unitary is locally equivalent to a controlled unitary in
Corollary \ref{cr:sl}. Next we show that $U$ can be implemented by
LOCC and a maximally entangled state $\ket{\Ps_r}={1\over \sqrt
r}\sum^r_{i=1}\ket{ii}$, where $r=\min\{d_A^2,d_B\}$ in Lemma
\ref{le:costSR3}. Next, we apply our result to solve a special case
of a conjecture on the ranks of multipartite quantum states, see
Conjecture \ref{cj:rank}.

Controlled unitary operators are one of the most easily accessible
and extensively studied quantum operators. For example, the
controlled NOT (CNOT) gate is essential to construct the universal
quantum two-qubit gate used in quantum computing \cite{bbc95}.
Experimental schemes of implementing the CNOT gates have also been
proposed, such as cavity QED technique \cite{sw95} and trapped ions
\cite{cz00}. Recently CNOT gates have been proved to be decomposed
in terms of a two-qubit entangled gate and single qubit phase gates,
which could be implemented by trapped ions controlled by fully
overlapping laser pulses \cite{msv14}. Next, multiqubit graph states
for one-way quantum computing are generated by a series of
controlled-Z gates \cite{br01}. Third, controlled phase gates have
been used to construct the mutually unbiased bases (MUBs)
\cite{wpz11} and graph states for which the violation of
multipartite Bell-type inequalities have been experimentally
demonstrated \cite{lzj14}. These applications (and those not
mentioned above) could be improved by the strengthened connection
between nonlocal and controlled unitaries presented in this paper.

The rest of this paper is organized as follows. In Sec.
\ref{sec:pre} we introduce the preliminary knowledge, and propose
Conjecture \ref{conj:control} as the main question in this paper. In
Sec. \ref{sec:conj} we prove Conjecture \ref{conj:control} when
$d_A=2,3$, and we propose its applications on general nonlocal
unitaries in Sec. \ref{sec:app}. Finally, we conclude in Sec.
\ref{sec:conclusion}.

\section{\label{sec:pre} Preliminaries}

Let $\cH=\cH_A\otimes\cH_B$ be the complex Hilbert space of a
finite-dimensional bipartite quantum system of Alice and Bob. We
denote by $d_A,d_B$ the dimension of $\cH_A$ and $\cH_B$,
respectively. It is known that $\cH$ is spanned by the computational
basis $\ket{i,j},i=1,\cdots,d_A$, $j=1,\cdots,d_B$. We shall denote
$I_k=\sum^k_{i=1} \proj{i}$. For convenience, we denote
$I_A=I_{d_A}$, $I_B=I_{d_B}$ and $I=I_{d_Ad_B}$ as the identity
operator on spaces $\cH_A,\cH_B$, and $\cH$, respectively. Let
$d=d_Ad_B$, and $U,V\in \Un(d)$ be two unitary matrices on the space
$\cH$. We say that $U,V$ are equivalent under stochastic local
operations, or \textit{SL-equivalent} when there are two locally
invertible matrices $S_1,S_2\in\GL(d_A)\times \GL(d_B)$ such that
$U=S_1VS_2$. In particular, we say that $U,V$ are \textit{locally
equivalent} when $S_1,S_2$ are unitary.

A unitary matrix $U$ on $\cH$ has Schmidt rank $n$ if there is a
decomposition $U=\sum^n_{j=1}A_j \ox B_j$ where the $d_A\times d_A$
matrices $A_1,\cdots,A_n$ are linearly independent, and the
$d_B\times d_B$ matrices $B_1,\cdots,B_n$ are linearly independent. 
Such decomposition will be called Schmidt expansion in this paper,
but note that the same term may sometimes imply the additional requirement that $A_j$ (and $B_j$)
are orthogonal under the Hilbert-Schmidt inner product.
We say that $U$ is a \textit{controlled unitary gate}, if $U$ is
locally equivalent to $\sum^{d_A}_{j=1}\proj{j}\ox U_j$ or
$\sum^{d_B}_{j=1}V_j \ox \proj{j}$. To be specific, $U$ is a
controlled unitary from the A or B side, respectively. Clearly the
matrices $U_j,V_j$ are unitary. We further say that system A (or B)
controls with $n$ terms if $U=\sum^n_{i=1} P_i \ox U_i$ (or
$\sum^n_{i=1} U_i \ox P_i$), where the $U_1,\cdots,U_n$ are unitaries 
and the $P_i$ are orthogonal projectors, i.e., $P_iP_j=\d_{ij}P_i$.

It is known that any multipartite (i.e., nonlocal) unitary gate of
Schmidt rank 2 is a controlled unitary \cite{cy13}. However there 
are bipartite unitaries of Schmidt rank 4, e.g. the two-qubit SWAP gate \cite{cy13},
that are not controlled unitaries.  It is then an interesting
question to characterize the bipartite unitaries of Schmidt rank 3.
Formally, we investigate the following conjecture in the next
section.
 \bcj
 \label{conj:control}
Any bipartite unitary operator of Schmidt rank 3 is a controlled
unitary operator.
 \ecj

To approach this conjecture, we generalize the concept of controlled
unitary gate, see an example in the next paragraph. We split the space into a direct sum:
$\cH_A=\op^m_{i=1} \cH_i$, $m>1$, $\dim\cH_i=m_i$, and
$\cH_i\perp\cH_j$ for distinct $i,j=1,\cdots,m$. We say that $U$ is
a \textit{block-controlled unitary (BCU) gate controlled from the A
side}, if $U$ is locally equivalent to $\sum^m_{i=1}
\sum^{m_i}_{j,k=1} \ketbra{u_{ij}}{u_{ik}}\ox V_{ijk}$ where
$\ket{u_{ij}}\in \cH_i$ for $j,k=1,\cdots,m_i$, and $m>1$. Note that the $V_{ijk}$ are not necessarily unitary. For simplicity we
denote the decomposition as $\op_A V_i$ where
$V_i=\sum^{m_i}_{j,k=1} \ketbra{u_{ij}}{u_{ik}}\ox U_{ijk}$, and we
denote $\abs{V_i}_A=m_i$. We have $UU^\dg=\sum^m_{i=1} P_{i} \ox
I_{B}=I$, where $P_i$ is the projector on the space $\cH_i$. So the
BCU from the A side can be understood as the direct sum of nonlocal
unitaries on the spaces $\cH_i\ox\cH_B$, $i=1,\cdots,m$. In
particular if $m_i=1$ for all $i$, then $U$ degenerates to a
controlled unitary gate from the A side. So a BCU has more general
properties than those a controlled unitary gate has. One may
similarly define the BCU gate controlled from the B side.

Although the controlled unitary gate is a BCU gate, the converse is
wrong. An example is the following qutrit-qubit unitary gate:
$U={1\over2}(I_2\ox
I_2+\s_x\ox\s_x+\s_y\ox\s_y+\s_z\ox\s_z)+\proj{3}\ox I_2$, where
$\s_x,\s_y,\s_z$ are the standard Pauli operators acting on the 2-dimensional subspace of $\cH_A$ spanned by $\ket{1},\ket{2}$. They can be regarded as qutrit operators by letting $\s_i\ket{3}=0$ for $i=x,y,z$. By definition $U$
is a BCU gate from $A$ side with $m=2$, $\cH_1=\lin\{\ket{1},\ket{2}\}$ and $\cH_2=\lin\{\ket{3}\}$. If $U$ is a controlled unitary from $A$
or $B$ side, then it has Schmidt rank at most 3 or 2. It is a
contradiction with the fact that $U$ has Schmidt rank 4. So $U$
is not a controlled unitary.

Since a nonlocal unitary and controlled unitary may be not locally
equivalent, one may ask when they are locally equivalent. The
question has been addressed in \cite[Lemma 2]{cy13}, which says the 
$A_j$ in the Schmidt expansion of $U$ have simultaneous singular value decomposition,
i.e. they are simultaneously diagonal under the same two local unitaries before and after them.
However the lemma is not very operational in practice. Below we present an
operational criterion based on \cite[Corollary 5]{mm11} and
\cite[Lemma 2]{cy13}.  Note that \cite{cs10} also cited some related
work of the authors of \cite{mm11}, and made implicit use of the same lemma below, in studying the entanglement
cost of more general types of bipartite unitaries.
 \bl
 \label{le:japan}
Let $U=\sum^r_{j=1} A_j \ox B_j$ be a nonlocal unitary of Schmidt
rank $r$. Then $U$ is a controlled unitary from the $A$ side if and only
if $A_i A_j^\dg$, $i,j=1, \cdots, r$ are all normal and commute with
each other, and $A_i^\dg A_j$, $i, j = 1, \cdots, r$ are all normal
and commute with each other.
 \el
 \bpf
The assertion immediately follows from \cite[Corollary 5]{mm11} and
\cite[Lemma 2]{cy13}.
 \epf

Note that the matrices $A_j$ (resp. $B_j$) are not necessarily orthogonal to each other.

\section{\label{sec:conj} Proving Conjecture \ref{conj:control} for $d_A=2,3$}

Conjecture \ref{conj:control} trivially holds for $d_A=1$. In this
section we show that Conjecture \ref{conj:control} holds when one of
the systems $A,B$ is a qubit or a qutrit. The first case is
demonstrated by the following theorem.
 \bt
 \label{thm:rk3da2}
Any bipartite unitary on $2\times d_B$ of Schmidt rank 3 is locally
equivalent to a controlled unitary controlled from the $B$ side.
 \et
 \bpf
Let $U$ be a bipartite unitary on $2\times d_B$ of Schmidt rank 3.
Suppose $U$ has an operator Schmidt expansion $U=\sum_{j=1}^3 E_j
\otimes F_j$. Using the orthogonality under the Hilbert-Schmidt
inner product, there is a $2\times2$ matrix $E_4$ orthogonal to
$E_1,E_2,E_3$. Let the nonnegative real numbers $a$, $b$ be the
singular values of $E_4$. Up to local unitaries we may assume
$E_4=a\proj{0}+b\proj{1}$. Since $U$ has Schmidt rank 3, $U$ is
locally equivalent to the unitary $U_1=\sum_{j=1}^3 A_j \otimes B_j$
where $A_1=\ketbra{0}{1}$, $A_2=\ketbra{1}{0}$,
$A_3=b\ketbra{0}{0}-a\ketbra{1}{1}$.

Let $\cH_B=\op^k_{i=1} V_i$ be an orthogonal decomposition and the
diagonal matrix $P_i$ the projector on the subspace $V_i$, $\forall
i$. So $P_iP_j=\d_{ij}P_i$ and $\sum^k_{i=1} P_i = I_B$. Up to local unitaries on $\cH_B$ we may assume the orthogonal decomposition
$B_3=\sum^k_{i=1} c_i P_i$, where the diagonal entries $c_i>c_j\ge0$ for all $i<j$. Since
$U_1$ is unitary, we have
 \bea\label{eq:I_B1}
 b^2 B_3^\dag B_3 + B_2^\dag B_2
 = B_1^\dag B_1 + a^2 B_3^\dag B_3
 =I_B,
 \\
 \label{eq:I_B2}
 b^2 B_3 B_3^\dag + B_1 B_1^\dag
 = B_2 B_2^\dag + a^2 B_3 B_3^\dag
 =I_B.
 \eea
Taking the trace in Eqs.~\eqref{eq:I_B1} and \eqref{eq:I_B2}, we
have $a=b>0$. Since $U_1$ is unitary, we have
 \bea
 \label{eq:I_B3}
B_3^\dag B_1 = B_2^\dag B_3, \quad\quad  B_1 B_3^\dag = B_3 B_2^\dag.
 \eea
Since $B_3=B_3^\dg=\sum^k_{i=1} c_i P_i$, it follows from \eqref{eq:I_B3} that $B_3^2 B_1 = B_1 B_3^2=B_3 B_2^\dag B_3$.
So $B_1$ commutes with $B_3^2$. One can similarly show that $B_2$ commutes with $B_3^2$. Since $c_i>c_j\ge0$ for all $i<j$, we have
$B_1=\op^k_{i=1} G_{i}$ and $B_2=\op^k_{i=1} H_{i}$, where the
square blocks $G_{i},H_{i}$ act on the space $V_i$, $\forall i$. By
\eqref{eq:I_B3} we have $G_1=H_1^\dg,\cdots,G_{k-1}=H_{k-1}^\dg$ and
$c_kG_k=c_kH_k^\dg$. It follows from \eqref{eq:I_B1} and
\eqref{eq:I_B2} that
 \bea
 \label{eq:normal}
B_1^\dag B_1=B_1 B_1^\dag=B_2^\dag B_2=B_2 B_2^\dag=I_B-a^2 B_3
B_3^\dag.
 \eea
So the matrices $G_1,\cdots,G_{k-1}$ and $H_1,\cdots,H_{k-1}$ are
normal. If $c_k>0$ then $B_1=B_2^\dg$ and $G_k$ is also normal by
\eqref{eq:normal}. So $B_1 $ is normal, and $B_1,B_2,B_3$ are
simultaneously diagonalizable under unitary similarity
transform. So $U_1$ is locally equivalent to a controlled
unitary controlled from the $B$ side. Since $U,U_1$ are locally
equivalent, the assertion follows. On the other hand if $c_k=0$, by
\eqref{eq:normal} we have $G_kG_k^\dg=H_kH_k^\dg=P_k$, i.e., both
$G_k,H_k$ are unitary. So $B_1,B_2,B_3$ are simultaneously locally
equivalent to diagonal matrices, and the assertion follows. This
completes the proof.
 \epf

No controlled unitary on $2\times d_B$ of Schmidt rank 3 can be
controlled from A side, otherwise the Schmidt rank would be
decreased. Below we construct a controlled unitary on $3\times 3$ of
Schmidt rank 3 which is not controlled from A side. Let
$U=\sum^3_{i=1} V_i\ox\proj{i}$, where $V_i=U_i \op \proj{3}$,
$i=1,2,3$ and the $U_i$ are linearly independent unitaries acting on the space ${\rm span}\{\ket{1},\ket{2}\}$. 
One can verify that $U$ is a controlled unitary
of Schmidt rank 3 controlled from B side. If it is also controlled
from A side, then the 3-dimensional subspace $H$ spanned by the
$V_i$ is also spanned by three matrices of rank one. This is a
contradiction with the fact that there is no matrix of rank one in
$H$. So $U$ is not controlled from A side. Next let $U'=\sum^3_{i=1}
V_i\ox P_i$ be a controlled unitary on $3\times d_B$ and $B$ control
with three terms. Using a similar argument above, we can show that
$U'$ is not controlled from $A$ side.

It is known that \cite[Theorem 6]{cy13} shows two facts. Any
bipartite unitary $U$ of Schmidt rank 2 (i) is controlled from both
A and B sides, and (ii) has at least one of the two systems $A,B$
controlling with two terms.  Can these two statements be generalized to unitaries of Schmidt rank 3?
The bipartite unitary in Theorem \ref{thm:rk3da2} and $U,U'$ in the last paragraph have
Schmidt rank 3 and violates statement (i). Next we show that statement (ii) cannot be generalized to that one side controls with three terms.
Consider the controlled unitary
\begin{equation}\label{eq:V324}
V=I_2\ox\proj{1}+\s_x\ox\proj{2}+\s_z\ox\proj{3}+{\s_x+\s_z \over \sqrt2}\ox\proj{4}
\end{equation}
of Schmidt rank 3 on $2\times4$ system. Evidently A side cannot control with
three terms. If B side controls with three terms, then $V$ is locally equivalent to
$V'=\sum^3_{i=1}U_i\ox P_i$ where $P_i$ are pairwise orthogonal
projectors.  In any expansion of the Schmidt-rank-3 unitary $V$ of the form $V=\sum_{j=1}^3 A_j\ox B_j$, the subspace span$\{B_j\}$ is well-defined in the sense that it is determined solely by $V$ and is independent of the form of the expansion (as long as the expansion has only 3 terms), so for this particular $V$ this subspace is the 3-dimensional subspace $S_1$ spanned by the matrices $\proj{1},\proj{2}+{1\over\sqrt2}\proj{4},\proj{3}+{1\over\sqrt2}\proj{4}$,
because we can choose $A_j$ to be $I_2$, $\s_x$ and $\s_z$.
The corresponding subspace for $V'$ is span$\{P_i\}$, which contains two linearly independent matrices of rank one.
As $V$ and $V'$ are locally equivalent, the subspace $S_1$ also contains two linearly independent matrices of rank one. This is impossible
and hence B side cannot control with three terms.  Therefore the statement (ii) cannot be directly generalized to the case of Schmidt rank 3.

It is known \cite[Proposition 3]{Nielsen03} that a two-qubit unitary cannot have Schmidt rank exactly 3. Theorem~\ref{thm:rk3da2} implies an alternative proof for this fact. If a two-qubit unitary has Schmidt rank 3, then from Theorem \ref{thm:rk3da2} it must be controlled from B side which is 2-dimensional, hence the unitary has Schmidt rank at most 2, a contradiction with the assumption that it has Schmidt rank 3.

In the paragraph below Proposition 3 in \cite{Nielsen03}, it was conjectured that a unitary operator on $d_A\times d_B$ system of Schmidt rank $k$ exists if and only if $k$ divides $d_Ad_B$. In the same paragraph there was an alternative conjecture. They are both false as the $V$ in Eq.~\eqref{eq:V324} is a counterexample.

To investigate Conjecture \ref{conj:control} with a qutrit system,
we present two preliminary lemmas.

 \bl
 \label{le:Ud=BCUd+1}
Assertion (i) below implies assertion (ii):
\\
(i) any bipartite unitary on $d_A \times d_B$ system of Schmidt rank 3 is
locally equivalent to a controlled unitary;
\\
(ii) any bipartite BCU (see the definition in the paragraph following Conjecture 1)  from the A side on $(d_A+1) \times d_B$ system of
Schmidt rank 3 is locally equivalent to a controlled unitary.
 \el
 \bpf
Let $U$ be a bipartite BCU from A side on $(d_A+1) \times d_B$ of
Schmidt rank 3. We may assume $U=U_1\op_A U_2$. Since $U$ has
Schmidt rank 3, $U_1,U_2$ have Schmidt rank at most 3. Since both
$\abs{U_1}_A,\abs{U_2}_A \le d_A$, it follows from (i) and
\cite{cy13} that both of them are equivalent to controlled unitaries.
If both of them are controlled from $A$ side, then (ii) holds. If
one of them is controlled from only $B$ side, then it has Schmidt
rank 3 \cite{cy13}.  Suppose it is $U_1$, then it has Schmidt expansion $U_1=\sum_{j=1}^3 A_j \ox B_j$,
where $A_j$ and $B_j$ are operators on $\cH_A$ and $\cH_B$, respectively.
As $U_1$  is controlled from B side, from \cite[Lemma 2]{cy13}, $B_j$ can be simultaneously diagonalized
under local unitaries.  Since $U$ also has Schmidt rank 3, it can be expanded 
in 3 terms with the same $B_j$ on the B side. Again using \cite[Lemma 2]{cy13}, $U$ is also a controlled unitary from $B$ side. Thus 
$(i)\ra(ii)$ holds.
 \epf

An open problem is whether the converse is true, i.e., $(ii)\ra(i)$.

 \bl
 \label{le:onezeroblock}
Any bipartite operator on $\cH$ of Schmidt rank at most $d_A$ is
locally equivalent to another operator $\sum^{d_A}_{i,j=1}
\ketbra{i}{j}\ox U_{ij}$ such that $U_{ij}=0$ for some pair of
subscripts $(i,j)$.
 \el
 \bpf
Let $V=\sum^{d_A}_{i,j=1} \ketbra{i}{j}\ox V_{ij}$ be an arbitrary
bipartite operator of Schmidt rank at most $d_A$. Using the row and
column operations, we only need to show that we can always realize
$V_{11}=0$. First this is evidently true when the blocks
$V_{i,1},\cdots,V_{i,d_A}$ are linearly dependent for some $i$. Next
suppose they are linearly independent for $i=1,\cdots,d_A$. Since
the Schmidt rank of $V$ is at most $d_A$, it becomes exactly $d_A$
now. So $V_{1,1},\cdots,V_{1,d_A}$ are in the $d_A$-dimensional
subspace spanned by $V_{2,1},\cdots,V_{2,d_A}$. There is a unit
vector $(x,y)$ such that the following $d_A$ matrix pencils
$xV_{1,1}+yV_{2,1},\cdots,xV_{1,d_A}+yV_{2,d_A}$ are linearly
dependent. Let $W$ be a unitary matrix of first row $(x,y,0,\cdots,0)$. Then the $d_A$ top blocks of size $d_B\times d_B$ in
$(W\ox I)V$ are exactly the above matrix pencils, so they are
linearly dependent. Now the claim follows from the first case. This
completes the proof.
 \epf

The assertion of this lemma can be easily generalized to the case in
which the bipartite operator $V$ is replaced by an isometry mapping
the space $\bC^n\ox\bC^q$ to $\bC^m\ox\bC^p$, i.e.,
$V=\sum^{m}_{i=1}\sum^{n}_{j=1} \ketbra{i}{j}\ox V_{ij}$ where
$V_{ij}$ is of size $p\times q$. Now we are in a position to prove
Conjecture \ref{conj:control} with $d_A=3$. We shall denote $A
\propto B$ for two linearly dependent matrices $A,B$.

  \bt
 \label{thm:3xd}
Any bipartite unitary on $3\times d_B$ of Schmidt rank 3 is locally
equivalent to a controlled unitary.
 \et
 \bpf
Let $U$ be a bipartite unitary on $3\times d_B$ of Schmidt rank 3.
For $d_B=2$ the assertion follows from Theorem \ref{thm:rk3da2}. 
The overall proof strategy is by induction over $d_B$. 
The inductive assumption is that the assertion holds for $d_B=2,\cdots,k-1$.
Then for $d_B=k$, we prove by considering the cases that $U$ is a BCU and not a BCU, respectively.

We claim that under the inductive assumption above, the assertion for $d_B=k$ holds for BCUs.
If $U$ is a BCU controlled from A side, then the claim follows from Lemma \ref{le:Ud=BCUd+1} and Theorem \ref{thm:rk3da2}. Let $U$ be
a BCU controlled from B side. We have $U=U_1 \op_B U_2$ where the
unitaries $U_i$ act on $\cH_A\ox\cH_i$, $i=1,2$ and
$\cH_1\op\cH_2=\cH_B$. Since $U$ has Schmidt rank 3, $U_1,U_2$ have
Schmidt rank at most 3. It follows from $U=U_1 \op_B U_2$ that both dimensions $\abs{U_1}_B, \abs{U_2}_B<k$. So
the induction hypothesis and \cite{cy13} (for Schmidt rank 3 and 2, respectively) imply that $U_1,U_2$ are
controlled unitaries. If both $U_1,U_2$ are controlled from the B
side then the claim follows. Suppose one of them, say $U_1$ is
controlled from the A side only. So $U_1$ has Schmidt rank 3
\cite{cy13}. Since $U$ also has Schmidt rank 3, it is a controlled
unitary from the A side (by the same argument as in the proof of Lemma~\ref{le:Ud=BCUd+1}), so the claim follows. 
From now on we assume that $U$ is not a BCU.

By Lemma \ref{le:onezeroblock} we may assume the bipartite unitary
$U=\sum^3_{i,j=1}\ketbra{i}{j} \ox U_{ij}$ with $U_{ij}$ of size
$d_B\times d_B$ and $U_{13}=0$. Since $U$ is unitary, the rows of submatrix $(U_{11},U_{12})$ are orthogonal to the rows of another submatrix
$\sum^3_{i=2}\sum^2_{j=1}\ketbra{i}{j} \ox U_{ij}$.
Since the former has rank $d_B$, the latter has rank at most
$d_B$. So the space spanned by the rows of two matrices $(U_{21},U_{22})$ and $(U_{31},U_{32})$ has dimension at most $d_B$. It implies that there is a unit vector $(x,y)$ such that the matrix pencil
$x(U_{21},U_{22})+y(U_{31},U_{32})$ has rank at most $d_B-1$. Let
$V_1$ be a $3\times3$ unitary with the bottom row $(0,x,y)$, and
$U'=(V_1 \ox I_B)U$. The bottom left $2d_B\times 2d_B$ submatrix of $U'$
consisting of four $d_B\times d_B$ blocks is exactly the above-mentioned
matrix pencil. So $U'$ is locally equivalent to
$W=\sum^3_{i,j=1}\ketbra{i}{j} \ox W_{ij}$, where the bottom row of
$W$ is $(0,\cdots,0,1)$. To obtain this form we have used $W=(I_A\ox S_B)U'(I_A\ox T_B)$, where the local unitaries $S_B$ and $T_B$ are for obtaining the first $2d_B$ and last $d_B$ elements of the bottom row, respectively.  As $W$ is unitary, $W_{33}$ is block-diagonal with a $(d_B-1)\times (d_B-1)$ block and a $1\times 1$ block.  Since $W$ is not a BCU, the three blocks $W_{31},W_{32},W_{33}$ are linearly dependent (otherwise all the 9 blocks $W_{ij}$ would be spanned by these three blocks and hence are block-diagonal, so $U$ is a BCU controlled from B side). This implies $W_{31}\propto W_{32}$. So $W$ is locally equivalent to
$W'=\sum^3_{i,j=1}\ketbra{i}{j} \ox W_{ij}'$ with
$W_{13}'=W_{31}'=0$. Since $W'$ is unitary, the rank of $W_{23}'$
and $W_{32}'$ are equal.  The first big row ($W'_{11},W'_{12},W'_{13}$) being orthogonal to the last big row
implies the sum of ranks of $W'_{12}$ and $W'_{32}$ is not greater than $d_B$.
Similarly, the sum of ranks of $W'_{21}$ and $W'_{23}$ is not greater than $d_B$.
Thus under local unitaries on $\cH_B$, $W'$ is equivalent to $V=\left(
                   \begin{array}{ccc}
                     v_1 & v_2 & 0 \\
                     v_3 & v_4 & v_5 \\
                     0 & v_6 & v_7 \\
                   \end{array}
                 \right)
$, where the $d_B\times d_B$ blocks $v_i$ have the expression $v_i =
\left(
   \begin{array}{cc}
     v_{i1} & v_{i2} \\
     v_{i3} & v_{i4} \\
   \end{array}
 \right)$, $i=1,\cdots,7$, and $v_{22}=v_{24}=v_{33}=v_{34}=v_{42}=v_{43}=v_{51}=v_{52}=v_{61}=v_{63}=0$, where
the equations $v_{42}=v_{43}=0$ are natural consequences of the other equations, and the choices of the local unitaries on $\cH_B$
before and after $W'$ are such that the other equations hold.
The blocks $v_{i1}$ are of size $(d_B-r) \times (d_B-r)$, $v_{i2}$
of size $(d_B-r) \times r$, $v_{i3}$ of size $r\times (d_B-r)$, and
$v_{i4}$ of size $r\times r$. Since $W$ is not a BCU, none of
$v_1,v_2,v_3,v_5,v_6,v_7$ is zero. So $r\in [1,d_B-1]$, and
$v_2,v_6$ are linearly independent. Let $H$ be the space spanned by
$v_2,v_3,v_5,v_6$. Since the Schmidt rank of $V$ is 3, we have $\dim
H = 2$ or $3$.

Suppose $\dim H =2$, so $v_3,v_5\in H=\lin\{v_2,v_6\}$. We have two
cases (1) $v_{23}=0$, $v_{64}\ne0$, and (2) $v_{23}\ne0$,
$v_{64}=0$. In case (1), we have $v_{32}=v_{53}=v_{62}=0$,
$v_{21}\propto v_{31}$ and $v_{54}\propto v_{64}$. If $v_4\in H$,
then $v_{21}$ and $v_{64}$ are both invertible. Since $V$ is
unitary, we have $v_{12}=v_{13}=v_{72}=v_{73}=0$. Then $V$ becomes a
BCU which gives us a contradiction with the assumption. So
$v_4\not\in H$. The space spanned by the $v_i$ is spanned by
$v_3,v_4,v_5$. Again $V$ becomes a BCU and we have a contradiction.
In case (2), we have $v_{31}=v_{21}=v_{54}=0$, $v_{32}\propto
v_{62}$ and $v_{23}\propto v_{53}$. If $v_4\in H$, then $v_4=0$.
Since $V$ is unitary, we have $r=d_B-r$ and $v_{23},v_{32}$ are
unitary, and the only nonzero blocks in $v_1$ and $v_7$ are $v_{11}$ and $v_{74}$, respectively.
So $V$ has Schmidt rank 4 and we have a contradiction. So $v_4\not\in H$. Since $V$ has Schmidt rank 3, we
have $v_{i3}\propto v_{23}$ and $v_{i2}\propto v_{32}$ ($i=1,\cdots,7$, same below). Since $V$ is
unitary, we have $v_{12}=v_{13}=v_{72}=v_{73}=0$. Since $V$ has
Schmidt rank 3, the three blocks $v_1,v_4,v_7$ are pairwise linearly
dependent. So $V$ is locally equivalent to a matrix $S$ the same as
$V$, except that $v_{11},v_{14}$ are replaced by scalar matrices, and hence
$v_{71},v_{74}$ are also scalar. It follows from $VV^\dg=I$ that $r=d_B-r$.  As $v_{23}\ne 0,v_{12}=v_{21}=0$,
and the rows of $V$ are normalized, we have that the singular values of $v_{11}$ are larger than those of $v_{14}$.  As the
blocks $v_1$ and $v_7$ are proportional, the singular values of $v_{71}$ are larger than those of $v_{74}$,
but this contradicts with that $v_{62}\ne 0,v_{64}=v_{73}=0$ and that the rows of $V$ are normalized.  Thus this case does not exist. So the case $\dim H =2$ is excluded.

Let $\dim H=3$. Up to local unitaries we may assume that $H$ is
spanned by $v_2,v_3,v_6$. Since $V$ has Schmidt rank 3, we have
$v_1,v_4,v_5,v_7 \in H$. Hence $v_{i3}\propto v_{23}$ and
$v_{i4}\propto v_{64}$.  We now prove $v_{23}\ne0$ and $v_{64}\ne0$.
Suppose $v_{23}=0$, then $v_{53}=v_{73}=0$, so that the submatrix formed by
$v_{44},v_{54},v_{64},v_{74}$ has rows orthogonal and normalized, so it is unitary
and hence its columns are normalized, therefore $v_{62}=v_{72}=0$.  Similarly
from $v_{13}=v_{23}=0$ we get that $v_{12}=v_{32}=0$.  So $U$ is a BCU controlled from the B side,
a contradiction.  So we have proved $v_{23}\ne0$.  Now suppose $v_{64}=0$, we have
$v_{44}=v_{54}=v_{74}=0$, so $v_{53}$ and $v_{73}$ are the only possibly nonzero blocks
on their respective rows, so they are nonzero, and $v_{i3}\propto v_{23}$ implies $v_{53}\propto v_{73}$, so the corresponding rows are not orthogonal,
violating that $V$ is unitary.  This proves $v_{64}\ne0$.

Next, we prove two statements: $v_{32}\propto v_{62}$, and $v_{21}\propto
v_{31}$. Suppose $v_{32},v_{62}$ are linearly independent. It follows from $v_4, v_5 \in H$ that $v_4=0$,
$v_{21}=v_{54}=0$, and $v_{23}\propto v_{53}$. Since $V$ is unitary,
by looking at the rows and columns that $v_{23}$ and $v_{53}$ are in, we have $r=d_B-r$, and hence $v_{13}=v_{14}=v_{71}=v_{73}=0$.
By $v_1\in H$ and $v_{32}\ne0$, $v_{64}\ne0$, we have $v_1\propto v_3$, and
as $v_{21}=v_{41}=0$, the rows containing $v_{11},v_{12}$ are proportional to
the rows containing $v_{31},v_{32}$, a contradiction with that $V$ is unitary.
Hence $v_{32}\propto v_{62}$. Next suppose $v_{21},v_{31}$ are linearly
independent. Since $ v_5 \in H$, we have $v_{53}=v_{62}=0$. 
And because $V$ is unitary, the submatrix formed by $v_{44},v_{54},v_{64},v_{74}$ is unitary, 
hence $v_{72}=v_{73}=0$. Since $v_7 \in H$, we have $v_{32}=0$.  From $v_1\in H$, we have $v_{12}=0$, hence $v_{14}$ is unitary,
which contradicts $v_{23}\ne0$. So $v_{21}\propto v_{31}$. 

From the results in the previous paragraph and that $v_i \in H$, we have $v_{i1}\propto v_{21}$, and $v_{i2}\propto v_{32}$. 
Therefore we may assume $v_{i1}=a_i A$, $v_{i2}=b_i B$, $v_{i3}=c_i C$, and $v_{i4}=d_i D$ with nonzero blocks $A,B,C,D$ for
$i=1,\cdots,7$. Since we have proved $v_{23}\ne0$, $v_{64}\ne0$, we have $c_2\ne0$, $d_6\ne0$. Since $V$ is unitary, we have
 \bea
 \label{eq:3xd1}
 (\abs{a_1}^2+\abs{a_2}^2) A A^\dg+\abs{b_1}^2 B B^\dg &=&
 I_{d_B-r},
 \\
 \label{eq:3xd2}
 (\abs{a_3}^2+\abs{a_4}^2) A A^\dg+\abs{b_3}^2 B B^\dg &=&
 I_{d_B-r},
 \\
 \label{eq:3xd3}
 (a_1a_3^*+a_2 a_4^*) A A^\dg+b_1b_3^* B B^\dg &=&
 0,
 \\
 \label{eq:3xd4}
 (\abs{c_1}^2+\abs{c_2}^2) C C^\dg+ \abs{d_1}^2 D D^\dg &=&
 I_{r},
 \\
 \label{eq:3xd5}
 \abs{c_5}^2 C C^\dg+ (\abs{d_4}^2+\abs{d_5}^2) D D^\dg &=&
 I_{r},
 \\
 \label{eq:3xd6}
 \abs{c_7}^2 C C^\dg+ (\abs{d_6}^2+\abs{d_7}^2) D D^\dg &=&
 I_{r},
 \\
 \label{eq:3xd7}
 c_5c_7^* C C^\dg+ (d_4d_6^*+d_5d_7^*) D D^\dg &=&
 0.
 \eea
Since $V$ is unitary and $c_2\ne0$, at least one of $b_1,b_3$ is
nonzero. If one of them is zero, then \eqref{eq:3xd1} and
\eqref{eq:3xd2} imply that $A$ is proportional to a unitary.  If
both $b_1,b_3$ are nonzero, then \eqref{eq:3xd1} and \eqref{eq:3xd3}
imply that $A$ is proportional to a unitary. So we have proved $A$
is always proportional to a unitary. It follows from \eqref{eq:3xd1}
and \eqref{eq:3xd2} that $B B^\dg$ is proportional to $I_{d_B-r}$.
Next, if one of $c_5,c_7$ is zero then \eqref{eq:3xd5} and
\eqref{eq:3xd6} implies that $D$ is proportional to a unitary. If
both $c_5,c_7$ are nonzero, then \eqref{eq:3xd6} and \eqref{eq:3xd7}
imply that $D$ is proportional to a unitary. So we have proved $D$
is always proportional to a unitary. It follows from $c_2\ne0$ and
\eqref{eq:3xd4} that $CC^\dg$ is proportional to $I_{r}$. So $V$ is
locally equivalent to the following matrix
 \bea\label{eq:vprime0}
 V'=
 \left(
   \begin{array}{cccccc}
     a_1 I_{d_B-r} & b_1 B & a_2 I_{d_B-r} & 0 & 0 & 0 \\
     c_1 C & d_1 I_{r} & c_2 C & 0 & 0 & 0 \\
     a_3 I_{d_B-r} & b_3 B & a_4 I_{d_B-r} & 0 & 0 & 0 \\
     0 & 0 & 0 & d_4 I_{r} & c_5 C & d_5 I_{r} \\
     0 & 0 & 0 & b_6 B & a_7 I_{d_B-r} & b_7 B \\
     0 & 0 & 0 & d_6 I_{r} & c_7 C & d_7 I_{r} \\
   \end{array}
 \right),
 \eea
where we still use the complex numbers $a_i,b_i,c_i,d_i$ and blocks
$B,C$ since there is no confusion. By adjusting the coefficients for the $B$ blocks, we may
assume that $BB^\dg=I_{d_B-r}$, and similarly we may assume $CC^\dg=I_r$. Since $V'$ is unitary, we have
 \bea
 \label{eq:v1}
(\abs{a_1}^2+\abs{a_2}^2)I_{d_B-r}+\abs{b_1}^2BB^\dg &=& I_{d_B-r},
 \\
 \label{eq:v2}
(\abs{a_3}^2+\abs{a_4}^2)I_{d_B-r}+\abs{b_3}^2BB^\dg &=& I_{d_B-r},
 \\
 \label{eq:v3}
\abs{d_1}^2I_r+(\abs{b_1}^2+\abs{b_3}^2)B^\dg B &=& I_{r}.
 \eea
Recall that one of $b_1,b_3$ is nonzero.  As $B^\dg B$ and $B B^\dg$ have the same rank, from the three equations above we have $d_B-r=r$. Hence $B$ and $C$ are square matrices and are unitaries.  Next we apply local unitaries to $V'$ to turn the $B$ into $I_r$, while preserving the other identity blocks in $V'$. The transform is given by $V''=(I_A\ox(I_r\oplus R))V'(I_A\ox(I_r \oplus R^\dg))$, where $I_r$ and $R$ are $r\times r$ unitaries acting on subspaces of $\cH_B$, and $R=B$.  So when $V''$ is expressed in the form of Eq.~\eqref{eq:vprime0}, the $B$ becomes $I_r$ while all the coefficients are unchanged.  The $C$ becomes a unitary matrix $C'=RC$ after the transform, and thus diagonalizable under a unitary similarity transform.  So the $C'$ blocks in $V''$ can be diagonalized with the following overall transform $X=(I_A\ox(S\oplus S))V''(I_A\ox(S^\dg \oplus S^\dg))$, where $S$ is a $r\times r$ unitary, and it can be verified that other blocks in $X$ are still $I_r$ with coefficients. So $V'$ is locally equivalent to a matrix $X$, which is still of the form \eqref{eq:vprime0} but $B$ and $C$ are replaced by diagonal matrices. So $X$ is a BCU from the B side, and we have a contradiction. This completes the proof.
 \epf

Let $U$ be a bipartite unitary on $d_A\times d_B$ of Schmidt rank 3
and $d_A=2,3$. It follows from Theorem \ref{thm:rk3da2} and
\ref{thm:3xd} that $U$ is a controlled unitary. We can further
decide the side from which $U$ is controlled by Lemma
\ref{le:japan}. To find out the explicit decomposition of $U$ into a
controlled unitary, we refer to an efficient algorithm constructed
in \cite{mm11} and references therein. The algorithm is proposed for
finding the finest simultaneous singular value decomposition for
simultaneous block diagonalization of square matrices under unitary
similarity.

\section{\label{sec:app} Characterization of nonlocal unitary operators}

In this section we propose a few applications of our results on
general nonlocal unitary operators. First we characterize the
equivalence of nonlocal unitaries and relate them to the controlled
unitaries. In Theorem \ref{thm:sllu} we show that the SL-equivalent
multipartite unitary operators are indeed locally equivalent. Using
it and Theorem \ref{thm:3xd} we can simplify the problem of deciding
the SL-equivalence of two bipartite unitaries of Schmidt rank 3 with
$d_A=2,3$. Using Theorem \ref{thm:sllu} we provide a sufficient
condition by which a bipartite unitary is locally equivalent to a
controlled unitary in Corollary \ref{cr:sl}. Next we propose an
upper bound on the quantum resources implementing bipartite
unitaries of Schmidt rank 3 with $d_A=2,3$, see Lemma
\ref{le:costSR3}. We also show that this upper bound is saturated
for some bipartite unitary. Third we apply our results to a special
case of Conjecture \ref{cj:rank} on the ranks of multipartite
quantum states. This conjecture is to construct inequalities
analogous to those in terms of von Neumann entropy such as the
strong subadditivity \cite{chl14}.

\subsection{Equivalence of nonlocal unitary operators}

We start by presenting the following observation on the
$SL$-equivalence of general nonlocal unitary operators.

 \bt
 \label{thm:sllu}
Suppose $U$ and $V$ are multipartite unitaries acting on $\cH_1
\otimes \cH_2 \otimes \cdots \otimes \cH_p$, and they satisfy
$U=(S_1\otimes S_2 \otimes \cdots \otimes S_p)V(T_1 \otimes T_2
\otimes \cdots \otimes T_p)$ for invertible operators $S_i$ and
$T_i$ acting on $\cH_1$, $\cH_2$, $\cdots$, $\cH_p$, respectively.
Then $U=(Q_1\otimes Q_2 \otimes \cdots \otimes Q_p)V(R_1 \otimes R_2
\otimes \cdots \otimes R_p)$, where $Q_i$ and $R_i$ are unitaries
acting on $\cH_1$, $\cH_2$, $\cdots$, $\cH_p$, respectively.  In
particular, when $S_i$ and $T_i$ are identity operators on any party
$i$, we can choose $Q_i$ and $R_i$ to be identity operators.
 \et

\bpf Suppose $S_i$ and $T_i$ have singular value decompositions of
the form $S_i=E_i C_i F_i$, $T_i=G_i D_i H_i$, where $E_i$, $F_i$,
$G_i$ and $H_i$ are unitaries, $C_i$ and $D_i$ are real diagonal
matrices with the diagonal elements sorted in non-increasing order.
The diagonal elements of $C_i$ and $D_i$ are the singular values of
$S_i$ and $T_i$, respectively.  Since $S_i$ and $T_i$ are
invertible, all the diagonal elements of $C_i$ and $D_i$ are
positive.

Let $U'=(E_1^\dag \otimes E_2^\dag \otimes \cdots \otimes E_p^\dag)
U (H_1^\dag \otimes H_2^\dag \otimes \cdots \otimes H_p^\dag)$, and
let $V'=(F_1 \otimes F_2 \otimes \cdots \otimes F_p) V (G_1 \otimes
G_2 \otimes \cdots \otimes G_p)$, then $U'$ and $V'$ are unitaries
and satisfy
\begin{equation}\label{eq:uprime}
U'=(C_1\otimes C_2 \otimes \cdots \otimes C_p) V' (D_A \otimes D_2
\otimes \cdots \otimes D_p).
\end{equation}
Using $U' U'^\dag=I$, where $I$ is the identity operator on the
entire space, we have
\begin{equation}
I=U' U'^\dag=(C_1\otimes C_2 \otimes \cdots \otimes C_p) V' (D_A^2
\otimes D_2^2 \otimes \cdots \otimes D_p^2) V'^\dag (C_1\otimes C_2
\otimes \cdots \otimes C_p),
\end{equation}
and using $V'^\dag = V'^{-1}$, we get
\begin{equation}\label{eq:vprime}
V' =(C_1^{2}\otimes C_2^{2} \otimes \cdots \otimes C_p^{2})  V'
(D_1^2 \otimes D_2^2 \otimes \cdots \otimes D_p^2).
\end{equation}
And since $C_i$ and $D_i$ are diagonal, $\tilde C := C_1^{2} \otimes
C_2^{2} \otimes \cdots \otimes C_p^{2}$ and $\tilde D := D_1^2
\otimes D_2^2 \otimes \cdots \otimes D_p^2$ are diagonal.  Consider
any nonzero element in the matrix $V'$, and let us suppose it is on
row $j$ and column $k$ of $V'$.  Then Eq.~\eqref{eq:vprime} implies
${\tilde C}_{jj} {\tilde D}_{kk}=1$, where ${\tilde C}_{jj}$ means
the $j$-th diagonal element of $\tilde C$, and ${\tilde D}_{kk}$ is
similarly defined.  And since $\tilde C$ and $\tilde D$ only contain
positive elements on their diagonals, we have ${\sqrt{\tilde
C}_{jj}} \sqrt{{\tilde D}_{kk}}=1$.  This holds for any 2-tuple
$(j,k)$ satisfying that the element on row $j$ and column $k$ of
$V'$ is nonzero, and since $C_1 \otimes C_2 \otimes \cdots \otimes
C_p$ and $D_1 \otimes D_2 \otimes \cdots \otimes D_p$ are diagonal,
this implies
\begin{equation}\label{eq:vprime2}
V' = (C_1 \otimes C_2 \otimes \cdots \otimes C_p) V' (D_1 \otimes
D_2 \otimes \cdots \otimes D_p).
\end{equation}
Together with Eq.~\eqref{eq:uprime}, we get $U'=V'$, hence
\begin{equation}\label{eq:uvnew}
U =(E_1 F_1 \otimes E_2 F_2 \otimes \cdots \otimes E_p F_p) V (G_1
H_1 \otimes G_2 H_2 \otimes \cdots \otimes G_p H_p),
\end{equation}
where $E_i F_i$ and $G_i H_i$ are unitaries by construction. From
the proof above we see that when $S_i$ and $T_i$ are identity
operators on any party $i$, we can choose $E_i$, $F_i$, $G_i$ and
$H_i$ to be identity operators. This completes the proof of
Theorem~\ref{thm:sllu}.
 \epf

The theorem implies that two SL-equivalent multipartite unitary
operators are indeed locally equivalent to each other.  Such two
unitaries can be viewed as the same nonlocal resource in quantum
information processing tasks. In contrast, two stochastic LOCC
(SLOCC)-equivalent pure states may be not locally equivalent, and
generally they can only probabilistically simulate each other in
quantum information processing tasks. For example, the 3-qubit W
state $\ket{\w}={1\over \sqrt3}(\ket{001}+\ket{010}+\ket{100})$
\cite{dvc00} and W-like state $\ket{\w'}={1\over 2}\ket{001}+{1\over
2}\ket{010}+{1\over \sqrt2}\ket{100}$ are SLOCC-equivalent but not
locally equivalent, as the bipartition of them gives rise to a
non-maximally entangled state and a maximally entangled state,
respectively.

It is known that the classification of multipartite states under
LOCC and SLOCC are different, because they are realized with
probability one and less than one, respectively. So the former is
more coarse-grained than the latter. For example, the three-qubit
pure states have infinitely many orbits under LOCC \cite{aacj00},
while there are only two kinds of fully entangled states under
SLOCC, namely the GHZ and W states \cite{dvc00}. In contrast,
Theorem \ref{thm:sllu} implies that the classification of
multipartite unitary operations under local unitaries and SL are
essentially the same, the latter does not give any additional
advantage the former does not have.  There are other ways of
classifying nonlocal unitaries, such as the LO, LOCC, and SLOCC
equivalences discussed in \cite{dc02}, which implicitly assume the
use of ancillas.

Based on the previous results we can simplify the decision of
SL-equivalence of two bipartite unitaries $U,V$ of Schmidt rank 3
and $d_A=2,3$. In practice this is motivated by the simulation of
one of them by the other, and the implementation of them. Using
Theorem \ref{thm:sllu} we only need to study the equivalence under
local unitaries. It follows from Theorem \ref{thm:3xd} that both
$U,V$ are controlled unitaries. They are not locally equivalent if
they are not controlled from the same side, which can be decided by
the algorithm in \cite{mm11}. Nevertheless, deciding the equivalence
of two controlled unitary controlled from the same side remains
unknown.

Below we characterize the controlled unitaries using Theorem
\ref{thm:sllu}.

 \bcr
 \label{cr:sl}
If a unitary $U$ on $\cH=\cH_A \otimes \cH_B$ is SL-equivalent to
\begin{equation}\label{eq:control_type}
V=\sum^{m}_{j=1} R_j \otimes V_j
\end{equation}
where $R_j$ are operators on $\cH_A$ satisfying
\begin{equation}
P_j R_j P_j=R_j,\quad\forall j,
\end{equation}
with $\{P_j\}$ being a set of mutually orthogonal projectors on
$\cH_A$, and $V_j$ are arbitrary operators on $\cH_B$. Then $U$ is
equivalent under local unitaries to the block diagonal form
\begin{equation}
U'=\sum_{j=1}^{m} P_j \otimes V'_j,
\end{equation}
where $V'_j$ are unitary operators on $\cH_B$.

In particular, if a unitary $U$ on $\cH$ is SL-equivalent to
$\sum^{d_A}_{j=1}\proj{j}\ox U_j$ for nonzero matrices $U_j$, then
$U$ is a controlled unitary gate controlled from the A side.
 \ecr
 \bpf
Note that the general case is reducible to the particular case by
first doing singular value decompositions of $R_j$, and at the end
noting that the final local unitaries $V'_j$ on $\cH_B$
corresponding to the same $R_j$ are the same. Hence we only need to
prove the particular case in the last paragraph of the assertion.

By hypothesis, $U$ is locally equivalent to a unitary
$W=\sum^{d_A}_{j=1}\ketbra{\a_j}{\b_j} \ox W_j$. The states
$\ket{\a_1},\cdots,\ket{\a_{d_A}}\in\cH_A$ are linearly independent,
and the states $\ket{\b_1},\cdots,\ket{\b_{d_A}}\in\cH_A$ are normalized (by absorbing constant factors into the corresponding $\ket{\alpha_j}$) and are also
linearly independent. Let $\ket{\g}\perp P$, and $P=I_{A}-\proj{\g}$
the projector on the hyperplane of $\cH_A$ spanned by
$\ket{\a_2},\cdots,\ket{\a_{d_A}}$. Since $W$ is unitary, we have
$\bra{\g}_A WW^\dg \ket{\g}_A = \abs{\braket{\g}{\a_1}}^2 W_{1}
W_{1}^\dg = I_{B}$. So the matrix $W_{1}$ is proportional to a
unitary matrix. We may replace $\ket{\a_2},\cdots,\ket{\a_{d_A}}$ by
any $d_A-1$ states of $\ket{\a_1},\cdots,\ket{\a_{d_A}}$ in the
above argument, and similarly obtain that the $W_i$'s are
proportional to unitary matrices, $i=2,\cdots,d_A$. So $U$ is
SL-equivalent to a controlled unitary from the $A$ side. The
assertion then follows from Theorem \ref{thm:sllu}. This completes
the proof.
 \epf

An explanation of Corollary~\ref{cr:sl} is as follows: if the effect
of a unitary is to stochastically implement a controlled type
operation of the form in Eq.~\eqref{eq:control_type}, then the
unitary must be a controlled unitary.

\subsection{Entanglement cost of implementing a bipartite unitary}

Computing the entanglement cost of implementing a nonlocal unitary
is an important question in quantum information \cite{dc02}.  For
this purpose a few protocols have been constructed. For example, one
can use teleportation \cite{bbc93} twice to implement a nonlocal
unitary by using LOCC and two maximally entangled states
$\ket{\Ps_{d_A}}$ ($d_A\le d_B)$, which contains $2\log_2 d_A$ ebits
\cite{ygc10}: Alice teleports her input system to Bob, and Bob does
the unitary locally, and teleports back the part of the output
system belonging to Alice to her. In ref.~\cite{ygc10}, another
protocol has been proposed to implement any bipartite controlled
unitary controlled from $A$ side by LOCC and the maximally entangled
state $\ket{\Ps_{d_A}}$. Using these protocols, and Theorem
\ref{thm:rk3da2} and \ref{thm:3xd}, we have
 \bl
 \label{le:costSR3}
Let $d_A=2,3\le d_B$. Any bipartite unitary of Schmidt rank 3 can be
implemented by using LOCC and the maximally entangled state
$\ket{\Ps_{k}}$, where $k=\min\{d_A^2,d_B\}$.
 \el
From this lemma, $\log_2 d_B$ ebits is an upper bound of the
amount of entanglement needed to implement a bipartite
unitary of Schmidt rank 3.  In the following we show that this upper
bound can be saturated for some unitary with $d_A=2$, $d_B=3$. Let
$U=I_2\ox \proj{1} + \s_x \ox \proj{2} + \s_y \ox \proj{3}$ be on the
space $\cH_A\ox\cH_B$, $A'$ be the ancilla qubit system, and the
bipartite space $\cK=\cH_{AA'}\ox\cH_{B}$. Let the product state
$\ket{\ps}={1\over\sqrt2}(\ket{11}+\ket{22})\ox{1\over\sqrt3}(\ket{1}+\ket{2}+\ket{3})\in
\cK$. Then $U\ket{\ps}\in\cK$ is a uniformly entangled state of Schmidt rank 3. That is, $U$ creates $\log_2 3$ ebits and therefore
implementing $U$ must cost at least so much entanglement
\cite{pv98}.  On the other hand, Lemma \ref{le:costSR3} implies that
$U$ can be implemented using $\log_2 3$ ebits and LOCC.

We leave as an open question whether there is a Schmidt-rank-3 unitary on $2\times 4$ system that needs more than $\log_2 3$ ebits to implement using LOCC.  Similar questions can be asked about Schmidt-rank-3 unitaries on $3\times d_B$ systems with $d_B\ge 4$.  This is a question about the lower bound of entanglement cost of unitaries, and there are a few results in the literature: Soeda \textit{et al} \cite{stm11} proved that one ebit of entanglement is needed for implementing a two-qubit controlled unitary by LOCC when the resource is a bipartite entangled state with Schmidt number two.  It is proved in Stahlke \textit{et al} \cite{sg11} that if the Schmidt rank of the resource state is equal to the Schmidt rank of the bipartite unitary, and the unitary can be implemented by the state using LOCC or separable operations, then the resource state must be uniformly entangled, i.e., with equal nonzero Schmidt coefficients; and higher Schmidt rank resource states may require less entanglement to implement the same unitary.  From these results we see that there are two branches to consider: using a resource state of Schmidt rank equal to that of the unitary or a state of higher Schmidt rank.

\subsection{A conjecture for the ranks of quantum states}

The following conjecture is proposed in \cite{chl14}.  In the
following $T$ denotes the matrix transpose.

 \bcj
 \label{cj:rank}
Let $R_1, \cdots, R_K$ be $m_1 \times n_1$ complex matrices, and let
$S_1, \cdots, S_K$ be $m_2 \times n_2$ complex matrices. Then
 \bea
 \label{eq:cjrank}
 \rank \bigg( \sum^K_{i=1} R_i \ox S_i^T \bigg)
 \le
 K \cdot \rank \bigg( \sum^K_{i=1} R_i \ox S_i \bigg).
 \eea
 \ecj
Note that $\rank(\sum^K_{i=1} R_i^T \ox S_i)=\rank(\sum^K_{i=1} R_i
\ox S_i^T)$ holds generally. The motivation of this conjecture is to
construct basic inequalities in terms of ranks of multipartite
quantum states, and some of them have been constructed in
\cite{chl14}. They are analogous to the inequalities in terms of von
Neumann entropy such as the strong subadditivity. Using the basic
inequalities one can constrain the relation of the ranks of
different marginals and quantify the multipartite entanglement
dimensionality.

The conjecture with $K=1$ is trivial, as the transpose does not
change the rank of a matrix. Next Conjecture \ref{cj:rank} with
$K=2$ has been proved in \cite{chl14}. However the conjecture with
$K\ge3$ is still an open problem and is considered to be highly
nontrivial in matrix theory. Nevertheless, the results in the last
section shed some light on the conjecture with $K=3$. Let
$U=\sum^3_{i=1} R_i \ox S_i$ be a $3\times d_B$ unitary matrix. Let
$U^\G=\sum^3_{i=1} R_i \ox S_i^T$ be the partial transpose of $U$
\cite{peres96} with the B side transposed. If $U$ is of Schmidt rank
3, Theorems \ref{thm:rk3da2} and \ref{thm:3xd} imply that $U$ is
locally equivalent to a controlled unitary; if the Schmidt rank of
$U$ is less than 3, $U$ is also locally equivalent to a controlled
unitary, according to \cite{cy13}.  The controlled unitary could be
controlled from either side, and in either case we have $\rank U^\G
= \rank U$.  Hence $\rank U^\G \le 3\cdot \rank U$, which is
Conjecture \ref{cj:rank} with $K=3$. Evidently, if Theorem
\ref{thm:3xd} can be generalized to any $d_A>3$, Conjecture
\ref{cj:rank} would hold for all Schmidt-rank-3 unitaries
$U=\sum^3_{i=1} R_i \ox S_i$.

\section{\label{sec:conclusion} Conclusions}

We have shown that the nonlocal unitary operator of Schmidt rank 3
on the $d_A \times d_B$ system is locally equivalent to a controlled
unitary when $d_A\le3$. Using this result we have shown that LOCC
and the $r\times r$ maximally entangled state of
$r=\min\{d_A^2,d_B\}$ are sufficient to implement such operators. We
also have shown that SL-equivalent nonlocal unitary operators are
indeed locally equivalent. In addition we have verified a special
case of Conjecture \ref{cj:rank} on the ranks of multipartite
quantum states, when the argument in the bracket of
\eqref{eq:cjrank} is a bipartite unitary of Schmidt rank 3 and
$d_A\le3$.

Unfortunately we are not able to prove Conjecture \ref{conj:control}
when $d_A>3$, as the proof of Theorem \ref{thm:3xd} cannot be easily
generalized. We believe that the generalization of this theorem will
prove Conjecture \ref{conj:control} and verify more cases of
Conjecture \ref{cj:rank}. Otherwise, the first counterexample to
Conjecture \ref{conj:control} might exist when $d_A=d_B=4$. The next
interesting question is to find generalizations of Lemma
\ref{le:Ud=BCUd+1}.  Finally, apart from the Schmidt rank, is there
another physical quantity which describes the local equivalence
between a nonlocal unitary and a controlled unitary? It remains to
investigate the connection between nonlocal and controlled unitaries
of arbitrary Schmidt rank.


\section*{Acknowledgments}

We thank Dan Stahlke and Scott Cohen for a careful reading and pointing out a few technical issues in an early version of the paper. We also thank Carlos A. Perez-Delgado, Joseph Fitzsimons and Masahito Hayashi for useful discussions. This material is based on research funded by the Singapore National Research Foundation under NRF Grant No. NRF-NRFF2013-01.

\bibliographystyle{unsrt}

\bibliography{controlunitary}

\begin{thebibliography}{10}

\bibitem{kmm13}
Vadym Kliuchnikov, Dmitri Maslov, and Michele Mosca.
\newblock {Asymptotically optimal approximation of single qubit unitaries by
  Clifford and T-circuits using a constant number of ancillary qubits}.
\newblock {\em Phys. Rev. Lett.}, 110:190502, May 2013.

\bibitem{cdgz13}
Lin Chen, Dragomir \ifmmode \check{Z}\else~\v{Z}.
  Djokovi\ifmmode~\acute{c}\else \'{c}\fi{}, Markus Grassl, and Bei Zeng.
\newblock Four-qubit pure states as fermionic states.
\newblock {\em Phys. Rev. A}, 88:052309, Nov 2013.

\bibitem{ccd13}
Lin Chen, Jianxin Chen, Dragomir~\v{Z}. Djokovi\'{c}, and Bei Zeng.
\newblock {Universal subspaces for local unitary groups of fermionic systems}.
\newblock March 2013.

\bibitem{cdg13}
Lin Chen, Dragomir~\v{Z}. Djokovi\'{c}, Markus Grassl, and Bei Zeng.
\newblock {Canonical form of three-fermion pure-states with six single particle
  states}.
\newblock June 2013.

\bibitem{ejp00}
J.~Eisert, K.~Jacobs, P.~Papadopoulos, and M.~B. Plenio.
\newblock Optimal local implementation of nonlocal quantum gates.
\newblock {\em Phys. Rev. A}, 62:052317, Oct 2000.

\bibitem{dc02}
W.~D\"{u}r and J.~I. Cirac.
\newblock {Equivalence classes of non-local unitary operations}.
\newblock {\em Quantum Information and Computation}, 2(3):240--254, 2002.

\bibitem{pv98}
Martin~B. Plenio and Vlatko Vedral.
\newblock Teleportation, entanglement and thermodynamics in the quantum world.
\newblock {\em Contemporary Physics}, 39(6):431--446, 1998.

\bibitem{ygc10}
Li~Yu, Robert~B. Griffiths, and Scott~M. Cohen.
\newblock Efficient implementation of bipartite nonlocal unitary gates using
  prior entanglement and classical communication.
\newblock {\em Phys. Rev. A}, 81:062315, Jun 2010.

\bibitem{bbc95}
Adriano Barenco, Charles~H. Bennett, Richard Cleve, David~P. DiVincenzo, Norman
  Margolus, Peter Shor, Tycho Sleator, John~A. Smolin, and Harald Weinfurter.
\newblock Elementary gates for quantum computation.
\newblock {\em Phys. Rev. A}, 52:3457--3467, Nov 1995.

\bibitem{cy13}
Scott~M. Cohen and Li~Yu.
\newblock {All unitaries having operator Schmidt rank 2 are controlled
  unitaries}.
\newblock {\em Phys. Rev. A}, 87:022329, Feb 2013.

\bibitem{sw95}
Tycho Sleator and Harald Weinfurter.
\newblock Realizable universal quantum logic gates.
\newblock {\em Phys. Rev. Lett.}, 74:4087--4090, May 1995.

\bibitem{cz00}
J.~I. {Cirac} and P.~{Zoller}.
\newblock {A scalable quantum computer with ions in an array of microtraps}.
\newblock {\em Nature}, 404:579--581, April 2000.

\bibitem{msv14}
Vladimir~S. Malinovsky, Ignacio~R. Sola, and Jiri Vala.
\newblock Phase-controlled two-qubit quantum gates.
\newblock {\em Phys. Rev. A}, 89:032301, Mar 2014.

\bibitem{br01}
Hans~J. Briegel and Robert Raussendorf.
\newblock Persistent entanglement in arrays of interacting particles.
\newblock {\em Phys. Rev. Lett.}, 86:910--913, Jan 2001.

\bibitem{wpz11}
M.~Wiesniak, T.~Paterek, and A.~Zeilinger.
\newblock Entanglement in mutually unbiased bases.
\newblock {\em New Journal of Physics}, 13(5):053047, 2011.

\bibitem{lzj14}
B.~P. Lanyon, M.~Zwerger, P.~Jurcevic, C.~Hempel, W.~D\"ur, H.~J. Briegel,
  R.~Blatt, and C.~F. Roos.
\newblock {Experimental violation of multipartite Bell inequalities with
  trapped ions}.
\newblock {\em Phys. Rev. Lett.}, 112:100403, Mar 2014.

\bibitem{mm11}
Takanori Maehara and Kazuo Murota.
\newblock Simultaneous singular value decomposition.
\newblock {\em Linear Algebra and its Applications}, 435(1):106 -- 116, 2011.

\bibitem{cs10}
Scott~M. Cohen.
\newblock Optimizing local protocols for implementing bipartite nonlocal
  unitary gates using prior entanglement and classical communication.
\newblock {\em Phys. Rev. A}, 81:062316, Jun 2010.

\bibitem{Nielsen03}
M.~A. Nielsen, C.~M. Dawson, J.~L. Dodd, A.~Gilchrist, D.~Mortimer, T.~J.
  Osborne, M.~J. Bremner, A.~W. Harrow, and A.~Hines.
\newblock Quantum dynamics as a physical resource.
\newblock {\em Phys. Rev. A}, 67:052301, May 2003.

\bibitem{chl14}
Josh Cadney, Marcus Huber, Noah Linden, and Andreas Winter.
\newblock Inequalities for the ranks of multipartite quantum states.
\newblock {\em Linear Algebra and its Applications}, 452(0):153 -- 171, 2014.

\bibitem{dvc00}
W.~D\"ur, G.~Vidal, and J.~I. Cirac.
\newblock Three qubits can be entangled in two inequivalent ways.
\newblock {\em Phys. Rev. A}, 62:062314, 2000.

\bibitem{aacj00}
A.~Acin, A.~Andrianov, L.~Costa, E.~Jane, J.~I. Latorre, and R.~Tarrach.
\newblock {Generalized Schmidt decomposition and classification of
  three-quantum-bit states}.
\newblock {\em Phys. Rev. Lett.}, 85:1560--1563, Aug 2000.

\bibitem{bbc93}
Charles~H. Bennett, Gilles Brassard, Claude Cr\'epeau, Richard Jozsa, Asher
  Peres, and William~K. Wootters.
\newblock Teleporting an unknown quantum state via dual classical and
  einstein-podolsky-rosen channels.
\newblock {\em Phys. Rev. Lett.}, 70:1895--1899, Mar 1993.

\bibitem{stm11}
Akihito Soeda, Peter~S. Turner, and Mio Murao.
\newblock Entanglement cost of implementing controlled-unitary operations.
\newblock {\em Phys. Rev. Lett.}, 107:180501, Oct 2011.

\bibitem{sg11}
Dan Stahlke and Robert~B. Griffiths.
\newblock Entanglement requirements for implementing bipartite unitary
  operations.
\newblock {\em Phys. Rev. A}, 84:032316, Sep 2011.

\bibitem{peres96}
Asher Peres.
\newblock Separability criterion for density matrices.
\newblock {\em Phys. Rev. Lett.}, 77:1413--1415, Aug 1996.

\end{thebibliography}

\end{document}